%% file: main.tex
\pdfoutput=1
\documentclass[journal]{IEEEtran}
\usepackage{amsmath,amsfonts}
\usepackage{adjustbox}
\usepackage{algorithmic}
\usepackage{algorithm}
\usepackage{array}
\usepackage[caption=false,font=normalsize,labelfont=sf,textfont=sf]{subfig}
\usepackage{textcomp}
\usepackage{stfloats}
\usepackage{url}
\usepackage{verbatim}
\usepackage{etoolbox}
\usepackage{listings}
\graphicspath{{./Figures/}}
\usepackage{import}
\usepackage{cite}
\usepackage[font=normalsize, labelfont=bf]{caption}
\usepackage{pgfplots}
\pgfplotsset{width=10cm,compat=1.9}
\usepackage{hyperref}
\hypersetup{
    colorlinks=true,
    linkcolor=black,
    citecolor=black,
    filecolor=magenta,
    urlcolor=cyan
    }

\usepgfplotslibrary{external}
\usepackage{tikz}
\usetikzlibrary{shapes.geometric, arrows}
\tikzstyle{startstop} = [rectangle, rounded corners, 
minimum width=3cm, 
minimum height=1cm,
text centered, 
draw=black, 
fill=red!30]
\tikzstyle{io} = [trapezium, 
trapezium stretches=true, % A later addition
trapezium left angle=70, 
trapezium right angle=110, 
minimum width=3cm, 
minimum height=1cm, text centered, 
draw=black, fill=blue!30]
\tikzstyle{process} = [rectangle, 
minimum width=3cm, 
minimum height=1cm, 
text centered, 
text width=3cm, 
draw=black, 
fill=orange!30]
\tikzstyle{decision} = [diamond, 
minimum width=3cm, 
minimum height=1cm, 
text centered, 
draw=black, 
fill=green!30]
\tikzstyle{arrow} = [thick,->,>=stealth]
\usepackage{tabularx,booktabs}
\newcolumntype{C}{>{\centering\arraybackslash}X} 
\begin{document}

\title{A Distributed Edge FLISR Solution \& Network Simulation Test Platform}

\author{Darren Leniston,~David~Ryan,~Ciaran~Malone,~and~Indrakshi~Dey,~\IEEEmembership{Senior Member,~IEEE}% <-this % stops a space
\thanks{D. Leniston, D. Ryan, C. Malone and I. Dey are with Walton Institute for Information and Communications Systems Sciences, Waterford, Ireland (Emails: darren.leniston@waltoninstitute.ie, david.ryan@waltoninstitute.ie, ciaran.malone@waltoninstitute.ie, indrakshi.dey@waltoninstitute.ie)}
\thanks{This material is based upon work supported by the {European Union’s Horizon 2020 research and innovation programme under the grant agreement No 883710}}
}

\markboth{Submitted to IEEE Transactions on Smart Grid vol.~xx no.~x}%
{Shell \MakeLowercase{\textit{et al.}}: A Sample Article Using IEEEtran.cls for IEEE Journals}


\maketitle
\import{Sections}{abstract.tex}

\import{Sections}{1_introduction.tex}

\import{Sections}{2_background.tex}

\import{Sections}{3_methods.tex}

\import{Sections}{4_experiments.tex}

\import{Sections}{5_results.tex}

\import{Sections}{6_conclusion.tex}

\bibliographystyle{IEEEtran}
\bibliography{main}

\end{document}

%% file: Sections/abstract.tex
\begin{abstract}
The energy sector is experiencing a paradigm shift with the swift adoption of distributed energy sources, renewables, electric vehicles, and an evolving consumer-utility relationship. This necessitates the strategic integration of advanced Information and Communication Technologies (ICT) and the Internet of Things (IoT) to address the emerging challenges. Grid resilience is paramount, as a dependable energy supply is the cornerstone of societal well-being and economic activity. The primary contribution of this research is to investigate the implementation of a novel grid resiliency strategy for the Irish context, employing Fault Location, Isolation and Service Restoration (FLISR) techniques in conjunction with Edge Computing. Through a comprehensive review of existing literature, original research activities, and meticulous data analysis, we aim to develop a solution that bolsters grid resilience and mitigates the impact of service disruptions for both consumers and utilities. Additionally, our work delves into the specific context of the Irish energy grid, including relevant policies and regulations, to ensure the proposed FLISR strategy is not only effective but also readily implementable.
\end{abstract}

%% file: Sections/1_introduction.tex
\section{Introduction}
\label{sec:introduction}

\IEEEPARstart{T}{\lowercase{he}} energy sector has undergone significant transformations, embracing renewable energy, electric vehicles, and software-driven solutions leveraging embedded devices. This shift towards a smart grid sets the stage for ongoing research, particularly in enhancing energy grid resilience. Building upon prior investigations within a European Horizon 2020 initiative, this study presents a pioneering automated grid resilience strategy tailored to the Irish energy grid context \cite{10.12688/openreseurope.14115.2}.

The energy grid is of vital importance in today's society and is a key piece of the infrastructure which enables activities across a range of industries, services and domestic settings. Like many aspects of the modern world, the energy industry has been experiencing a paradigm shift in the last number of years with the introduction of state-of-the-art Information and Communication \& Internet of Things technologies in an effort to modernise the energy sector; in terms of production, transmission and delivery to consumers \cite{6099519}. These new technologies are taking the form of smart sensors and
power monitoring at both the grid and domestic level, unlocking new data and granting greater insight into the state of the grid \cite{7123563}.

The research pursued in this document aims to meet the requirement for a resilient energy grid, utilising modern technologies and techniques, to reduce and mitigate the effects of disruptive events on the grid and their impact on the consumer and utility. To meet this goal, the research leverages the concept of Fault Location, Isolation and Service Restoration (FLISR), a grid resilience strategy which in the event of a fault on the grid, aims to automate the identification of said fault, isolating physically damaged sections of the network and restoring service to consumers on lines affected by loss of service but are otherwise undamaged.

In combination with the FLISR technique, the application of Edge Computing is explored in tandem with a novel representation of grid topology data to realise a distributed, fault-tolerant FLISR solution which is not fully reliant on a centralized cloud deployment, by moving subsets of the grid topology and FLISR logic to low-cost edge devices installed on existing grid infrastructure. In order to test, evaluate and demonstrate the solution, a simulation platform was developed. This research is framed in the context of the Irish distribution grid, in particular three distribution grids of  with differing topologies and features located throughout the country.

%\subsection{Motivation}
%\label{sec:motivation}

Recent times have seen an increase in the number and severity of weather events due to climate change, which cause wide-scale damage to infrastructure. Such events often lead to loss of energy supply to consumers, businesses and other institutions through damage caused to wires, transformers, substations, equipment and other grid assets \cite{7091066}. In relation to the Irish energy network, weather events such as ex-hurricane Ophelia in 2017 resulted in 5,500 damaged overhead lines and 385,000 homes \& businesses without power \cite{dso-annual-performance-report_2017}.

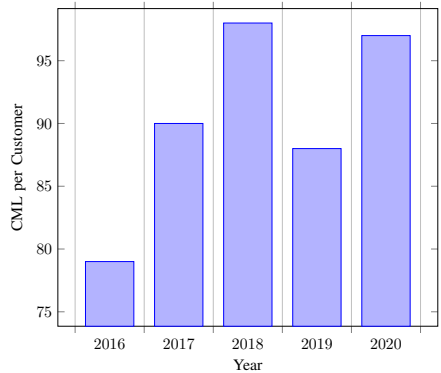
\begin{figure}[ht]
    \centering
    \begin{tikzpicture}[scale=0.6]
    \begin{axis}[
        x tick label style={/pgf/number format/1000 sep=},
        xlabel=Year,
    	ylabel=CML per Customer,
    	enlargelimits=0.05,
        ybar interval=0.7
    ]
    \addplot 
    	coordinates {(2020, 97) (2019, 88) (2018, 98) (2017, 90) (2016, 79) (2015, 75)};
    \end{axis}
    \end{tikzpicture}
    \caption{\label{fig:cml}CML figures 2016--2020}
\end{figure}

In addition, the Distribution System Operator (DSO) in Ireland is faced with fines by the regulator based on specific Key Performance Indicators (KPI) which are directly influenced by the amount of unplanned service interruptions and loss of supply, namely Customer Interruptions (CI) \& Customer Minutes Lost (CML). For reference, ex-hurricane Ophelia resulted in a total of 665.6 million CML, culminating in a 4.86 million euro fine imposed by the Irish energy regulator for that year \cite{dso-annual-performance-report_2017}.
However, 2017 was not an outlier in terms of total CML per customer, as detailed in the DSOs 2020 report \cite{networks_2021} and presented in \hyperref[fig:cml]{Figure \ref{fig:cml}}.

Techniques such as FLISR offer the possibility to drastically mitigate the effects of inclement weather events. However, there are limitations to the DSO in regards to deploying a FLISR solution to the grid, such as the cost of specialised equipment, in addition to the requirement for large-scale software platforms such as Advanced Distribution Management Systems (ADMS). Weather events can also affect communication networks, thereby reducing or even negating the benefit of a cloud-based or centralized FLISR implementation. With this factor in mind, the concept of edge computing is one that may greatly benefit the operation of a FLISR algorithm in the field, by moving a subset of the business logic and data processing away from a centralized system to the edge of the network. This would enable sections of the network to operate independently if needed to isolate and restore fault events, providing a fault-tolerant FLISR solution.

% subsection motivation (end)

%\subsection{Contribution}
%\label{sec:objectives}

The key objectives which framed the direction of the research activities centered around delivering a novel FLISR solution which can operate both centrally and in a distributed fashion at the network edge, and to test, evaluate and demonstrate the operation and effectiveness of the solution through the use of a simulation platform. The primary focus of this paper is to gauge how the developed FLISR solution may reduce CML in real-world operation following a framework;
\begin{itemize}
  \item \textbf{Effectiveness of the FLISR Solution} : We will evaluate the impact of the FLISR solution on critical DSO metrics used to measure fault handling performance. By comparing FLISR's performance to existing solutions or alternative FLISR approaches from other studies, we can assess its effectiveness in reducing fault duration, frequency, and other relevant KPIs. Additionally, we will consider the generalizability of these findings to the Irish energy grid by testing the solution on various grid topologies.
  \item \textbf{Centralized vs. Distributed FLISR Algorithms} : We will compare the performance of the centralized and distributed FLISR algorithms. A key aspect of the distributed approach is its fault tolerance, allowing it to function even when communication with the central algorithm is disrupted. However, this local operation comes with a limited view of the entire network. We will investigate whether the distributed algorithms can produce results comparable to the centralized approach despite this limitation.
  \item \textbf{FLISR Solution Adaptability} : We will assess the adaptability of the FLISR solution to different grid configurations. An effective FLISR solution should be able to handle a wide range of network topologies, including variations in structure, number of nodes, and the types of assets present within the grid. By testing the solution on diverse grids, we will evaluate its robustness and generalizability across the Irish energy infrastructure.
\end{itemize}
% subsection objectives (end)
% section introduction (end)

%% file: Sections/2_background.tex
\section{Related Works}
\label{sec:background}

%\subsection{Digitalisation of the Energy Grid}
%\label{sec:smart_grid}

A wide range of literature explores the concept of the smart grid and what the digitalisation and modernisation of the energy network means for the energy industry.
\textit{Fang et al.} \cite{6099519} describe a smart grid as utilising two-way flows of electricity and data to create an automated and distributed advanced energy delivery network, and through the application of ICT lead to a more efficient and flexible grid. The article describes the requirements for the smart grid from the perspective of three major systems; \textit{smart infrastructure systems}, \textit{smart management systems} and \textit{smart protection systems}. Of relevance are smart infrastructure systems, described as supporting two-way data flows via wired and wireless communications, and smart protection systems of which failure diagnosis and grid self-healing are a key factor, which directly relates to the requirements of the FLISR concept.

The application of ICT within the energy grid through the systems described above has lead to the realisation of the smart grid as outlined in an article by \textit{Güngör et al.} \cite{6011696} which describes the smart grid as a ``modern electric power grid infrastructure for improved efficiency, reliability and safety, with smooth integration of renewable and alternative energy sources, through automated control and modern communication technologies''. They describe how the reliability of energy infrastructure and supply can be supported and maintained in the smart grid, and how the impact of equipment failure, capacity constraints and natural accidents, can be mitigated by online power system monitoring, diagnostics and protection.

The smart grid's success relies heavily on the ability to exchange data reliably between smart devices in the field, such as those located within homes and businesses. This two-way flow of information is crucial for monitoring grid health, optimizing energy use, and enabling real-time decision-making. Therefore, the communication methods employed by the smart grid are of paramount importance. \emph{Güngör et al.} \cite{5406152} delve into how wireless communication methods can effectively address the challenges that arise during the transition to a smart grid. Traditionally, wired communication via cabling strung from pole to pole was the primary method for power system monitoring. This approach suffers from a significant drawback: high cost. The financial burden associated with wired infrastructure has limited its widespread implementation. \emph{Priya \& Malhotra} \cite{9231264} propose an alternative solution that leverages the capabilities of the existing telecommunication network to achieve efficient and reliable communication. They specifically focus on the potential of 5G technology, highlighting its strengths – namely, its high speeds and low latency – as ideal characteristics for capturing the vast amounts of data generated by smart grid devices in the field. This real-time data acquisition is essential for enabling rapid, automated control services that can manage, balance, and ensure the resilience of the entire grid.
% subsection smart_grid (end)

%\subsection{Fault Location, Isolation \& Service Restoration}
%\label{sec:flisr}

FLISR is a key technique for improving the reliability and resilience of the energy network – a critical requirement for Distribution Systems Operators (DSOs) and a frequent topic within smart grid research. \emph{Agüero} \cite{6344960} defines FLISR as a self-healing approach that leverages Distribution Automation (DA) technologies for remote monitoring, coordination, and operation of network assets across substation, feeder, and customer levels. By utilizing remotely controlled on-pole switching devices, FLISR can sectionalize faults and restore power automatically, minimizing the need for manual intervention by DSO staff. This automation promises significant improvements in reliability metrics and reduces costs associated with outages.

However, Agüero's research suggests that self-healing via FLISR is better suited for urban and suburban areas, where complex networks are common, rather than rural locations with predominantly radial distribution (single power sources). One might argue that FLISR-based resilience strategies are even more essential in these rural areas, where faults are often more frequent due to greater exposure to external factors. The US Department of Energy's 2014 report \cite {us_dept_energy} supports the effectiveness of FLISR. Analyzing five separate FLISR deployments across the US over a year-long period, the report highlights a significant decrease in both customers affected by service loss (270,000) and Customer Minutes Lost (CML) (38 million). This equates to a 45\% reduction in interruptions and a 51\% decrease in CML per fault. The report emphasizes that robust communication networks are vital for successful FLISR operations. These networks facilitate monitoring and remote control of switching devices and require interoperable interfaces to ensure seamless integration with the DSO's diverse grid systems.

FLISR implementations vary widely in their devices, communication methods, automation levels, and deployment locations. For example, CenterPoint's ``Self-Healing Grid" relies on Intelligent Grid Switching Devices but requires manual validation by an operator. In contrast, Duke's ``Self-Healing Teams" utilizes reclosers, breakers, and sensors for a fully automated solution. FLISR trials highlight key lessons learned. Telecommunication networks supporting FLISR need to be even more resilient than the power systems they manage. Utilities must also plan carefully for the additional steps involved in successful FLISR deployment, including equipment installation, updates, and staff training. Interestingly, Georgia Power found that developing a simulator for FLISR testing and validation reduced the need for field trials and aided in training operators.

Eriksson et al. \cite{7000578} discuss the limitations of centralized FLISR systems, such as the risk of single-point failure and the challenges of handling vast amounts of real-time data on large network topologies. They propose a distributed FLISR approach utilizing a Multi-Agent System (MAS) tested on a radial network typical of rural areas. Employing Prim's Algorithm for FLISR logic, their solution demonstrates effective grid stabilization and fault restoration within a cyber-physical testbed. Of particular relevance is the report by Ireland's ESB Networks, an output of the H2020 SOGNO project \cite{sogno_report}. Their FLISR implementation uses Load Break Fault Make (LBFM) circuit breakers for fault detection. Currently, manual interventions often lead to extended restoration times and higher CML. While the SOGNO project's FLISR solution shows promise, ESB Networks highlights its reliance on a centralized instance and telecommunications as a significant limitation.
% subsection flisr (end)

%\subsection{Edge Computing}
%\label{sec:edge_computing}

Cloud computing has revolutionized large-scale data collection and processing, but the rise of IoT demands a shift. Shi et al. \cite{7488250} term this the "post-cloud era" where embedded IoT devices not only generate vast amounts of data but also process and act upon it at the network's edge. This minimizes the strain on cloud computing services and network bandwidth. Edge computing brings computational activities closer to the data sources, enabling devices to consume and produce data.

Successful edge computing solutions require careful deployment and management of edge resources. Taleb et al. \cite{7931566} emphasize the role of orchestration frameworks in integrating, managing, and updating edge devices. Virtualization, such as containerized software, facilitates the delivery and maintenance of the software services that run on field-deployed edge devices. Edge computing has significant potential for smart grids. Zhang et al. \cite{9492236} suggest it can address challenges like high computational costs and transmission delays in smart grid operations. Similarly, Chen et al. \cite{8727940} highlight edge computing's benefits in improving grid resiliency and self-healing. By reducing dependency on centralized cloud systems, edge computing enables more efficient control, monitoring, and quicker local decision-making. Additionally, edge devices' proximity to grid assets improves latency and provides the DSO with fine-grained network visibility, especially at the MV and LV levels.
% subsection edge computing (end)

%\subsection{Summary}
%\label{sec:summary}

Extensive research highlights the challenges and pathways to achieving a resilient smart grid. FLISR (Fault Location, Isolation, and Service Restoration) emerges as a powerful tool for improving grid resilience, with various implementations explored in the literature (algorithms, communication protocols, levels of automation). However, a key concern is ensuring that FLISR solutions are themselves fault-tolerant, avoiding reliance on centralized systems or core telecommunication networks. Distributed FLISR appears to offer a solution to this challenge.

Edge computing presents an ideal paradigm for implementing distributed FLISR. By decentralizing data processing and decision-making from cloud-based systems, edge computing promises reduced latency and decreased dependency on core systems for distributed FLISR. Research further suggests that edge computing is critical for the broader realization of the smart grid due to the massive data generation and computational demands involved. Additionally, the cost and complexity of modifying legacy grid infrastructure make edge computing even more attractive. 

% subsection summary (end)
% section background (end)

%% file: Sections/3_methods.tex
\section{Methods}
\label{sec:methods}

\subsection{Analysis \& Import of Grid Network Topology}
\label{sec:analysis_import}

As the underlying grid topologies support both the FLISR algorithms and the simulation platform, extracting this information from the available data served as a starting point for development. The data is derived from a Geographical Information System (GIS) which maps the distribution network through the use of geospatial data in addition to metadata related to each individual asset, whether the entity is a pole, switch, substation or other asset type. The properties contained depend on the asset type, including connected poles, and phases (RST) which specifies if a pole is three phase (MV) or single phase (LV). Additional properties include which poles are equipped with Load Break Fault Make (LBFM) or Recloser circuit breakers and which of these are configured as ``Normally Open'' and therefore serve as the boundary between network lines. The work of transforming this data consisted of four primary steps; mapping the base GIS data to a graph model, analysing the key properties to include in the final dataset, determining a method of approximating the number of connected customers to each pole and correlating the identified properties to a consistent model.

% subsection analysis_import (end)

\subsubsection{Mapping of Relational Data to Graph Database Format}
\label{sec:mapping}

\begin{figure}[ht]
  \centering
  \includegraphics[width=0.5\textwidth]{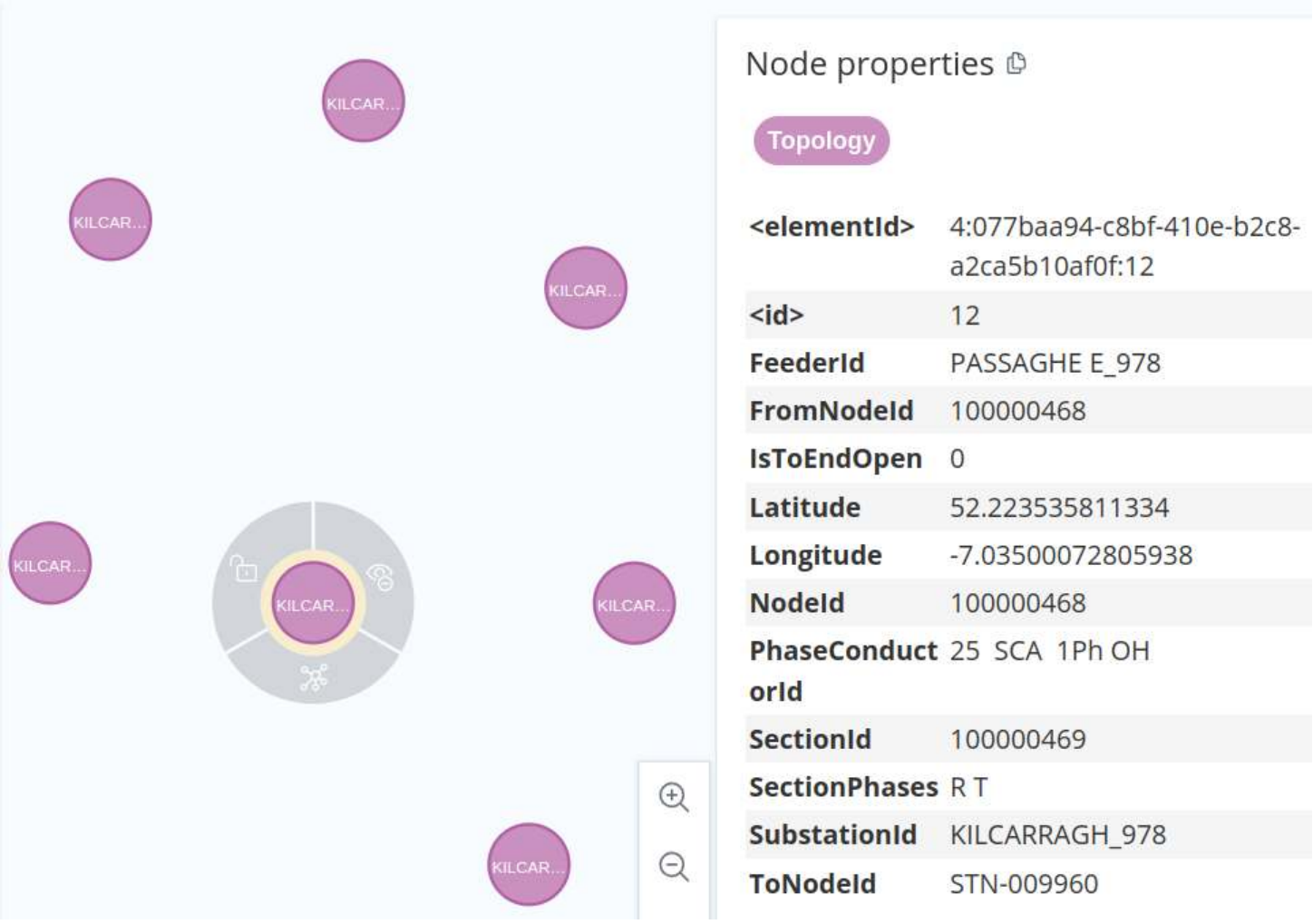}
  \caption{\label{fig:graph_database}Topology data ingested in graph database}
\end{figure}

The original GIS dataset represents the distribution network topology in a relational model. However, the inherent interconnectedness of power grids lends itself more naturally to a graph representation. We employ Graph Databases for their ontological view of relationships \cite{8509057}, allowing us to represent the network as a connected graph $G(V, E)$, achieving several benefits: a) Nodes ($V$):  In our graph model, nodes represent individual assets of the distribution network. This includes poles, transformers, substations, feeder lines, switches and other relevant entities. Each node is associated with properties extracted from the relational GIS dataset. b) Edges ($E$):  Edges in our graph represent the physical connections between network assets. These edges can be directed or undirected, depending on the nature of the connection (e.g., power flow direction). The graph database is employed utilizing ``neo4j'' graph as illustrated in \hyperref[fig:graph_database]{Fig.~\ref{fig:graph_database}}.

% subsubsection mapping (end)

\subsubsection{Identification of Key Node Properties}
\label{sec:identification}

The GIS dataset contains fifty tables, and was explored thoroughly to discern useful data considering relevant assets and to organise it in such a way that it could be efficiently loaded into the simulation platform and ingested by the FLISR algorithms.

\begin{figure}[ht]
  \centering
  \includegraphics[width=0.32\textwidth]{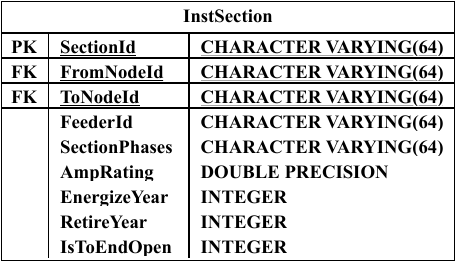}
  \caption{\label{fig:inst_section}InstSection Schema}
\end{figure}

A key table identified ``InstSections'' forms the hub of the data schema and maps to a majority of other entities via the property ``SectionId''. In the context of the GIS system individual poles are referred to as ``nodes'', and therefore the properties ``FromNodeId'' and ``ToNodeId'' determine how poles are connected, additionally the property ``SectionPhases'' specifies whether a section is part of the MV or LV network, \hyperref[fig:inst_section]{Fig.~\ref{fig:inst_section}} details a subset of the ``InstSection'' schema.

\begin{figure}[ht]
  \centering
  \includegraphics[width=0.3\textwidth]{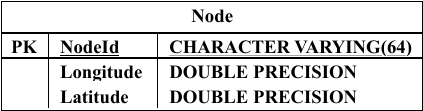}
  \caption{\label{fig:node_schema}Node Schema}
\end{figure}

Geolocation data for nodes is found in the ``Nodes'' table, in Geographic (latitude \& longitude) coordinate format as shown in \hyperref[fig:node_schema]{Figure \ref{fig:node_schema}}.

\begin{figure}[ht]
  \centering
  \includegraphics[width=0.3\textwidth]{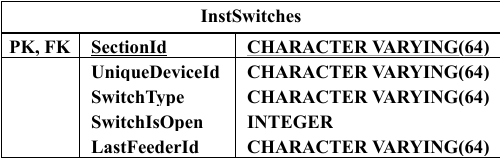}
  \caption{\label{fig:switch_schema}InstSwitches Schema}
\end{figure}

Information on nodes with on pole equipment, are contained within the ``InstSwitches'' table, which includes a number of properties as illustrated in \hyperref[fig:switch_schema]{Fig.~\ref{fig:switch_schema}}, most notably ``SectionId'' and ``UniqueDeviceId'' for linking to the ``InstSection'' and ``Node'' entities, in addition to ``SwitchType'' where LBFM circuit breakers are denoted as ``Soules'' and ``SwitchIsOpen'' which specifies whether the asset is configured to be a normally open point.

\begin{figure}[ht]
  \centering
  \includegraphics[width=0.35\textwidth]{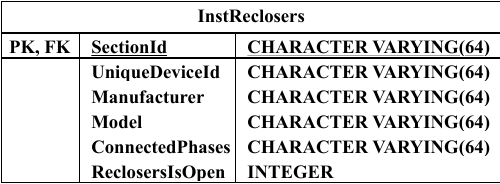}
  \caption{\label{fig:recloser_schema}InstReclosers Schema}
\end{figure}

While Reclosers are denoted as circuit breakers, within the GIS dataset these are denoted as an entity named ``InstReclosers'' with a number of properties as described in \hyperref[fig:recloser_schema]{Figure \ref{fig:recloser_schema}}. Similar to the previous entity, the ``InstReclosers'' table contains the ``SectionId'' and ``UniqueDeviceId'' properties. Where it differs is the inclusion of ``Manufacturer'' and ``Model'', the assets of interest are denoted as ``Nu-Lec'' model ``N Series''. Reclosers configured as normally open are specified by the ``RecloserIsOpen'' property.

% subsubsection identification (end)

\subsubsection{Estimating connected customers at each node}
\label{sec:estimation}

\begin{figure}[ht]
  \centering
  \includegraphics[width=0.35\textwidth]{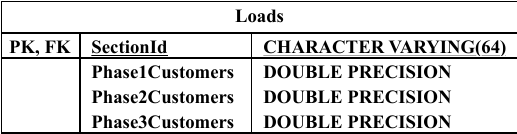}
  \caption{\label{fig:loads_schema}Loads Schema}
\end{figure}

A key requirement of the simulation platform is to provide metrics based on the outputs of the FLISR algorithms. As the GIS data is anonymised, it does not contain fine-grained customer related data, therefore it was necessary to derive some approximation of the amount of connected customers by identifying properties that may aid in determining this data point. A table was identified named ``Loads'', described in \hyperref[fig:loads_schema]{Figure \ref{fig:loads_schema}} which references connected customers on each phase and links to the ``InstSection'' table via the
``SectionId'' property. We introduce some formal notation to define our customer estimation method: $L$ is the set of all load records in the Loads table. Each load record $l$ in $L$  contains attributes section ID and number of customers per phase. $C_{\phi, i}$ is the set of customers connected to a specific phase ($\phi \in {RST}$) within a network section $i$ and $S_i$ is the set of total customers within network section $i$. To estimate the total customers in a section, we take the union of customers across all phases which can be mathematically represented as;
\begin{align}
    {S_i =  \bigcup\limits_{\phi \in {RST}} C_{\phi, i}}
\end{align}
This means that the total customers in a section $i$ is the combined count of customers connected to its  $R$-phase, $S$-phase, and $T$-phase.

 By correlating and appending the ``Phase$<$N$>$Customers'' to each ``InstSection'' it is possible to see an approximation of customers per section of network. The word ``approximation'' is important to note here however, as specific sections of network had a large number of customers, while viewing these locations via satellite perspective showed a low number of buildings in the area, one possibility is that these may be industrial, farming or other large customers which are treated differently within the GIS system. Accounting for these outliers, a total number of customers was added as a property to on pole circuit breaker assets in the data, to allow for the collection of customer and CML related metrics.

% subsubsection estimation (end)

\subsubsection{Definition of node model}
\label{sec:model}

For the Irish network, assets such as on pole circuit breakers are identified with a unique SCADA identifier, LBFM switches are denoted with an ``S'' identifier e.g., S514, with Reclosers denoted with an ``R'' e.g., R341. Assets that communicate with the DSO SCADA system are given an additional identifier which maps to their SCADA identification, named an ASDU (Application Service Data Unit). The relevant properties ascertained from the entities included in the GIS data, supplemented with the approximated customers and SCADA specific identifiers, formed the properties of each object in the graph database.
 
Here we formulate a robust framework for Fault Detection for the FLISR platform; Let $S$ be the set of all intelligent switching devices (ISDs) on the distribution grid, also known as Fault Line Current Limiters (FLCLs) or reclosers. These devices are equipped with Fault Passage Indicators (FPIs) and Loss of Voltage Indicators (LVIs) to enable fault detection and isolation. Let, $s \in S$ be an individual switch (ISD) belonging to the set of switches $S$. Each switch $s$ is characterized by its current operational state (open or closed), its configuration (normally open or normally closed), and the status of its FPI and LVI indicators. $FPI(s)$ is a Boolean function that returns True if the Fault Passage Indicator has been triggered on switch $s$, indicating that the current flowing through the switch has exceeded a predefined threshold, suggesting a fault on the line segment controlled by the switch. Otherwise, the function returns False. $LVI(s)$ is a Boolean function that returns True if the Loss of Voltage Indicator has been triggered on switch $s$, signifying a voltage drop below a certain threshold on the line segment controlled by the switch. This could be caused by a fault or other grid disturbances. Otherwise, the function returns False. $NO(s)$ is a Boolean function that returns True if the switch $s$ is configured as a normally open point. The function returns False if the switch is configured as normally closed.

Using this notation, we can express the decision rules more formally:
\begin{itemize}
    \item \textbf{Rule 1}: Isolation of Faulted Section: This rule prioritizes isolating the faulted section of the grid. If both the FPI and LVI are triggered on a switch $s$, it signifies a high probability of a fault on the line segment controlled by that switch. In this scenario, the decision rule instructs the system to open the switch $s$ (Open($s$)), effectively isolating the fault and preventing further propagation of the fault current. This can be formally expressed as:
    \begin{align}
        \forall s \in S, (FPI(s) \land LVI(s)) \implies Open(s)
    \end{align}
    
    \item \textbf{Rule 2}: Maintaining Power Delivery:  The next rule ensures continuity of power supply to healthy sections.  If a Loss of Voltage (LVI) is detected on a switch configured as a normally closed point ($\neg NO(s)$), it indicates a potential fault downstream. However, opening the switch in this case could unnecessarily isolate healthy loads. Therefore, the rule dictates that the switch s should remain closed (Closed(s)) to maintain power delivery to healthy sections. This is formally expressed as:
     \begin{align}
        \forall s \in S, (LVI(s) \land \neg NO(s)) \implies Closed(s)
    \end{align}
    
    \item \textbf{Rule 3}: Recloser Operation:  The final rule addresses normally open switches (NO(s)). If a Loss of Voltage (LVI) is detected on a normally open switch, the rule instructs the system to close the switch (Close(s)) in anticipation of a potential transient fault clearing. This is formally expressed as:
         \begin{align}
            \forall s \in S, (LVI(s) \land NO(s)) \implies Close(s)
    \end{align}
\end{itemize}
% subsubsection model (end)

\subsection{Network Simulation Platform}
\label{sec:sim_platform}

\begin{figure}[ht]
  \centering
  \includegraphics[width=0.5\textwidth]{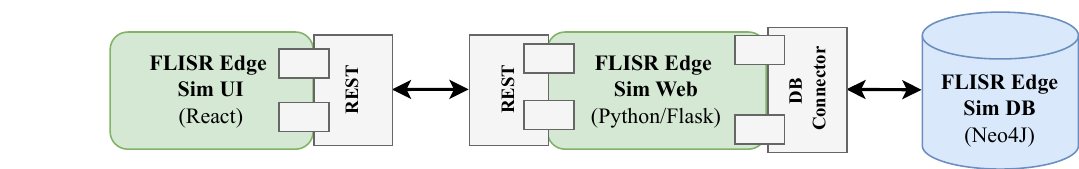}
  \caption{\label{fig:sim_arch}FLISR Edge simulation platform architecture}
\end{figure}

 The FLISR Edge simulation platform enables the simulation of fault events serves to both test and evaluate the FLISR algorithms in lieu of actual hardware, these simulations are based on the distribution grid topologies obtained from the GIS dataset. The simulation platform leverages a micro-service architecture, in addition these services are hosted within a containerised environment to enable simple deployment on either local systems or remote host \cite{7436659}. \hyperref[fig:sim_arch]{Figure \ref{fig:sim_arch}} describes the architecture implemented for the simulation platform.

% subsection sim_platform (end)

\subsection{Overview of Trial Site Topologies}
\label{sec:trial_features}

 The distribution grids investigated throughout the development of the FLISR Edge solution, are dispersed throughout Ireland, located in County Waterford, County Kerry and County Westmeath and this section will describe these trial sites in terms of their structure.

\subsubsection{Trial site 1 topology}
\label{sec:waterford_trial}
\begin{figure}[ht]
  \centering
  \includegraphics[width=0.38\textwidth]{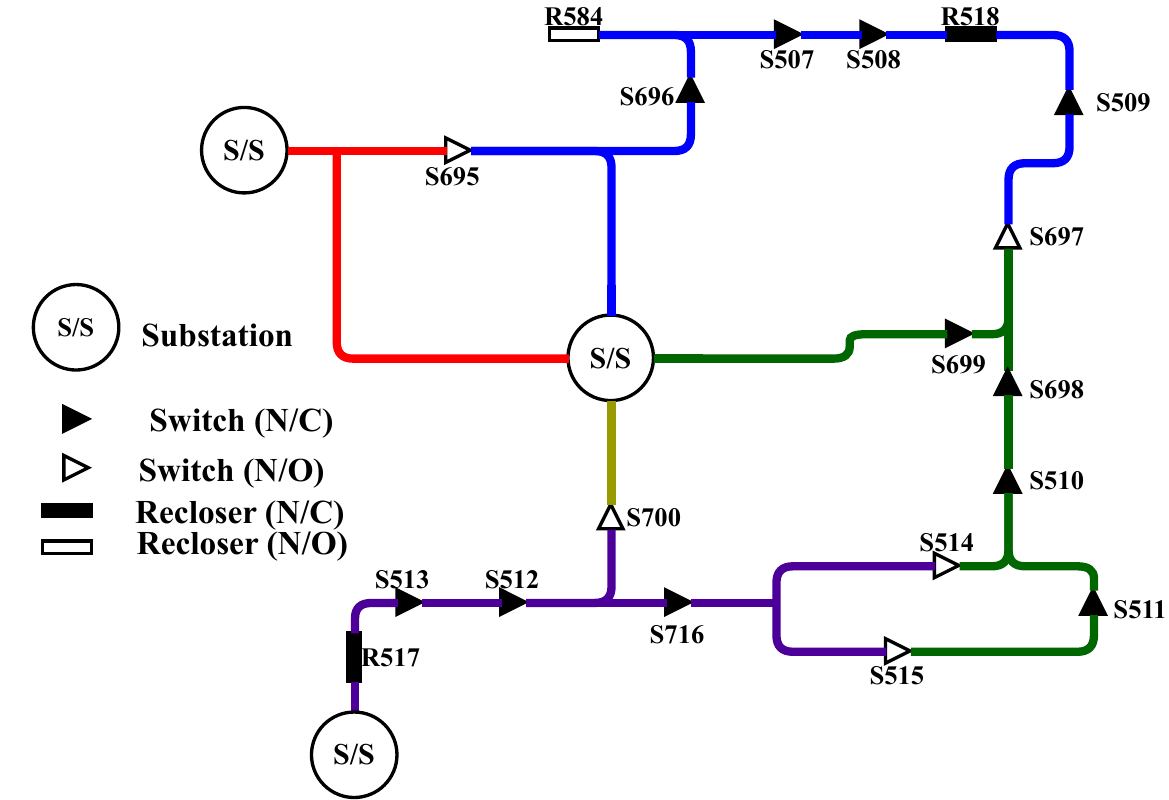}
  \caption{\label{fig:waterford_site}Waterford trial site circuitry}
\end{figure}

The distribution grid located in Waterford is complex, consisting of an intricate mesh network as presented in \hyperref[fig:waterford_site]{Figure \ref{fig:waterford_site}}. This network is fed by three substations, with a total of eighteen on pole circuit breakers installed, consisting of fifteen LBFM switches and three Reclosers. Taking into account the sites complexity and proximity to the coastline, there is a higher likelihood of fault events due to inclement weather and as such would be suitable candidate for testing a FLISR solution in the field.

% subsubsection waterford_trial (end)

\subsubsection{Trial site 2 topology}
\label{sec:kerry_trial}

\begin{figure}[ht]
  \centering
  \includegraphics[width=0.5\textwidth]{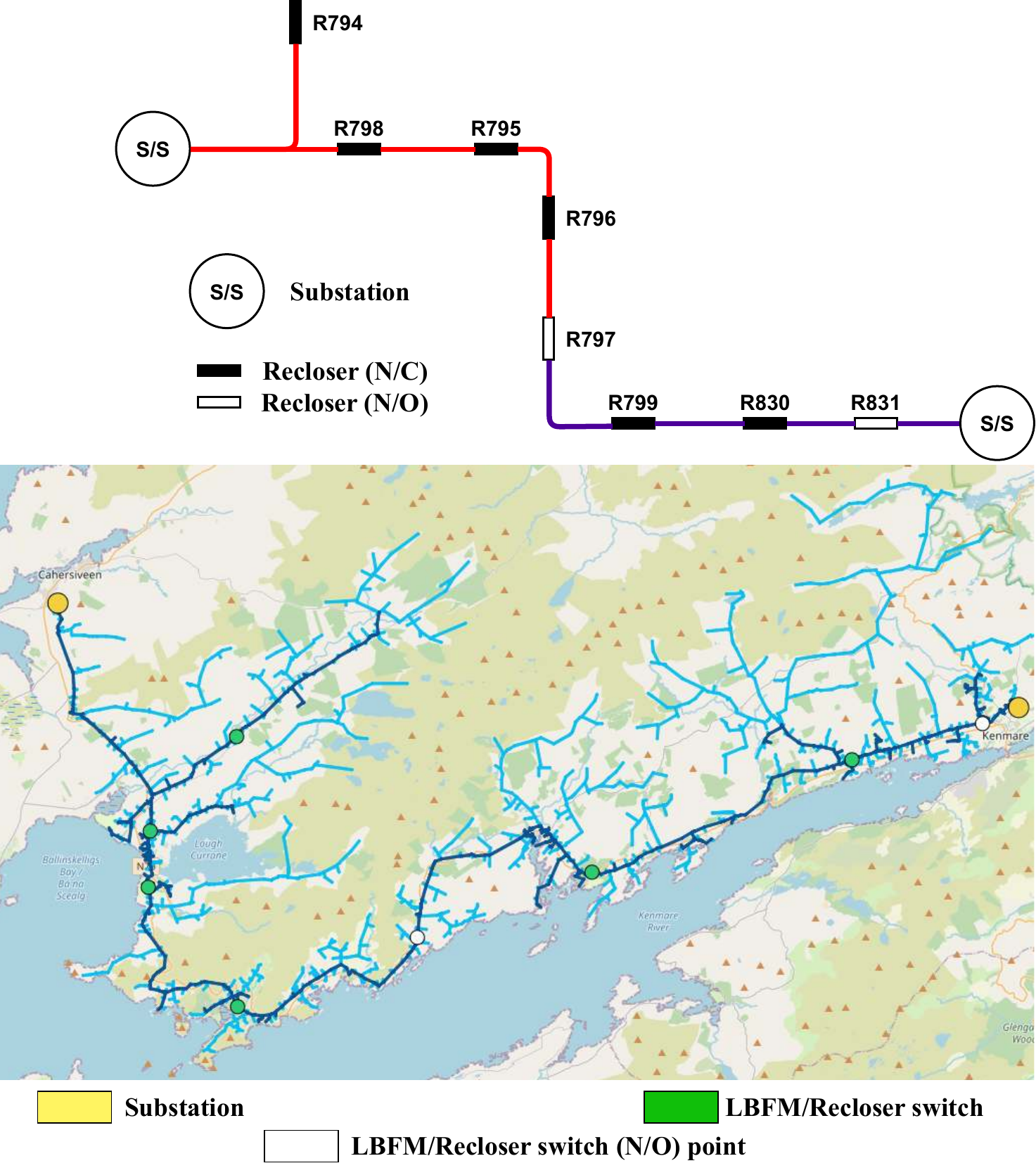}
  \caption{\label{fig:kerry_site}Kerry trial site circuitry}
\end{figure}

The distribution grid located in Kerry is situated by the coastline, fed at both ends via substation, with a total of eight Recloser switches, the circuitry and visualisation of this network is illustrated in \hyperref[fig:kerry_site]{Figure \ref{fig:kerry_site}}. Due to the structure of this network, there is a narrow scope to back feed supply in the event of a fault, and with its remote nature it is likely that network connectivity is limited, particularly in higher areas and mountainous valleys, making it a suitable use case for installation of edge devices to provide distributed FLISR functionality.

% subsubsection kerry_trial(end)

\subsubsection{Trial site 3 topology}
\label{sec:mullingar_trial}

\begin{figure}[ht]
  \centering
  \includegraphics[width=0.35\textwidth]{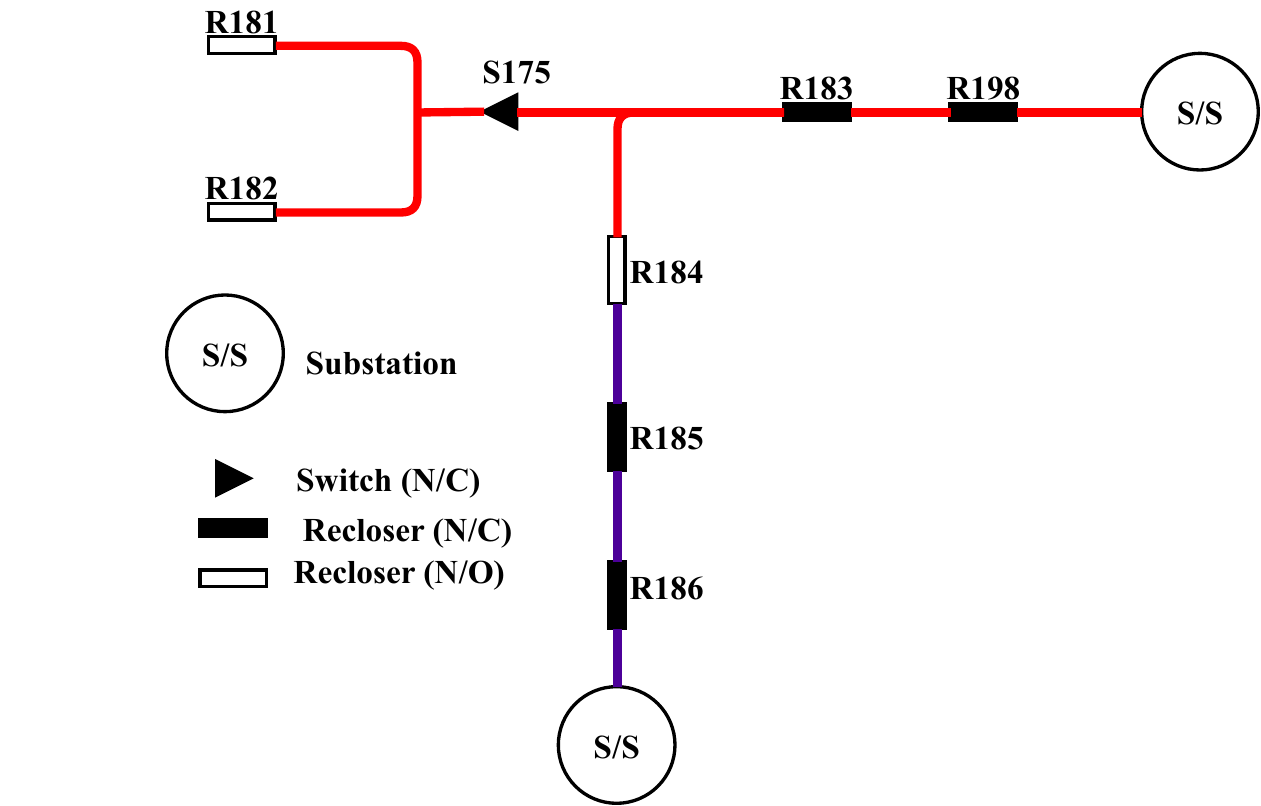}
  \caption{\label{fig:mullingar_site}Mullingar trial site circuitry}
\end{figure}

The distribution network located in Mullingar consists of two substations, and seven Recloser switches installed, \hyperref[fig:mullingar_site]{Figure \ref{fig:mullingar_site}} presents the circuitry and visualisation of the site. Two of the Reclosers are located at the boundary of the network and serve as normally open points, providing opportunity for back feeding energy supply from adjoining networks. Its location in the midlands likely leads to relatively strong network coverage and a number of fault events that are more representative of the country at large and therefore would serve as an ideal location to test the operation of a FLISR solution in a less unpredictable network.

% subsubsection mullingar_trial(end)
% subsection trial_features (end)

\subsection{FLISR Edge solution}
\label{sec:flisr_edge}

 The FLISR Edge solution consists of two applications of the FLISR concept, one of which is a centralized implementation deployed in a remote location, such as the DSOs own systems, with the second deployed on LBFM and Recloser assets by equipping them with an additional edge device running the distributed algorithm, which can act independently if necessary. This builds on the previous related research by expanding the capabilities of the FLISR algorithm, introducing bi-directional communication to directly control assets. In contrast, the previous iteration was limited to a simple generated email which was forwarded to DSO personnel \cite{10.12688/openreseurope.14115.2}.

The additional edge device will require compatible interfaces to link with existing equipment on the pole, sufficient processing capability to run the application and cellular connectivity to process data and control signals. With the rising popularity of low-cost and Edge computing, coupled with the growing rate of processing available to even low-power chips, the options available to provision a low-cost Edge computing device is much higher than in the previous decade \cite{8016573}. When considering a network such as Waterford, the cost of retrofitting existing LBFM switches with Reclosers to enable loop automation, costs 61,900 euros per unit according to ESB Networks 2023 standard prices report \cite{Networks_2023}. In comparison, exploring the application of low-cost Edge computing devices to enable loop automation on existing assets appears more financially feasible.

Assets equipped with switching hardware consist of a number of modules which allow for interfacing with the DSOs SCADA system for data comms and enacting control operations on the physical circuit breaker. At the core of this architecture is a Remote Terminal Unit (RTU), which streams real-time status data to and from the connected circuit breaker \cite{6032722}. Communication with the DSOs SCADA system conforms to the IEC  60870-5-101/104 protocol, based on TCP/IP communications and enables basic control tasks between centralized control centers and connected assets, through connection from the RTU to a IEC 101 Master \& IEC 104 Server \cite{6672100}. It is this IEC 104 server which provides the network communications with the DSO SCADA system, as well as a number of interfaces to connect external devices, including serial and Ethernet connectivity. It is these physical ports that will enable connection with the edge device, allow for the capture and parsing of the switch status messages without interfering with the existing stream to the DSOs SCADA system.

\begin{figure}[ht]
  \centering
  \includegraphics[width=0.4\textwidth]
  {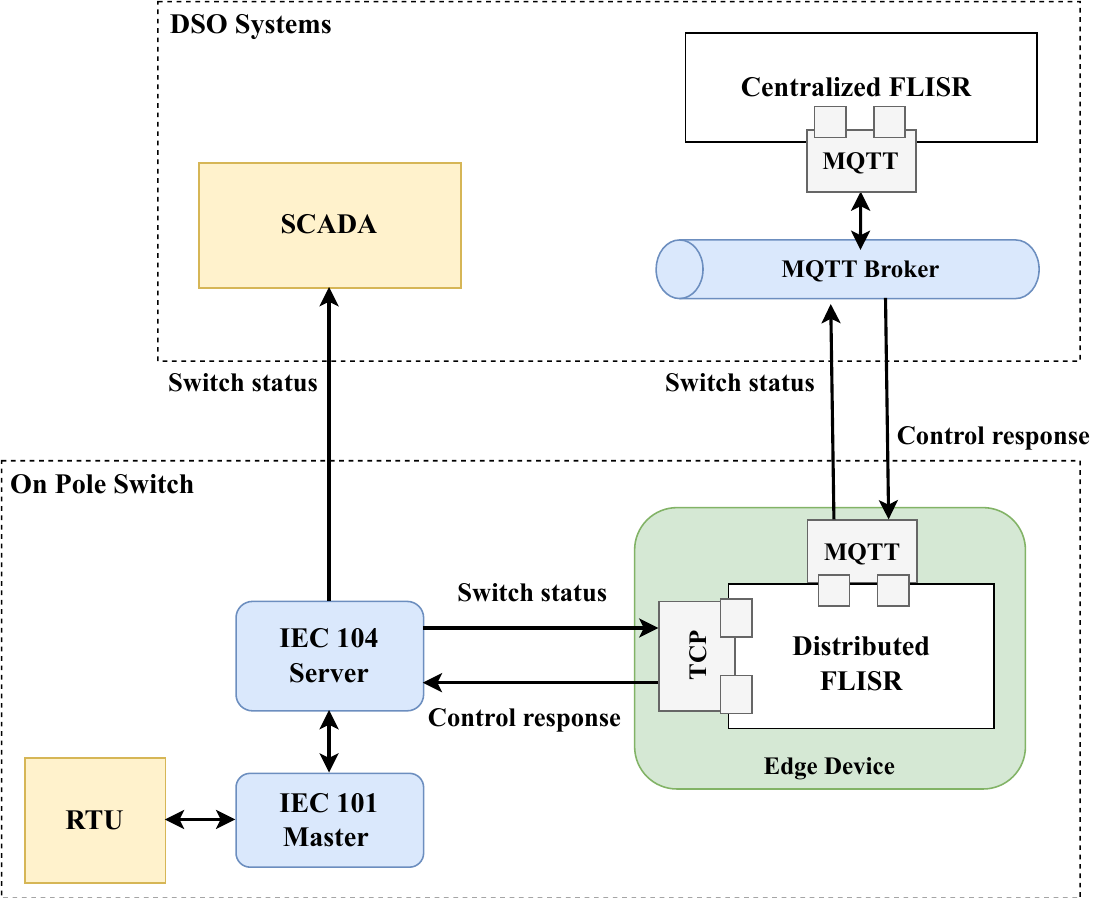}
  \caption{\label{fig:rtu_interaction}Edge device interfaces \& interaction}
\end{figure}

Within the context of the FLISR Edge algorithms operation, the edge device will act as a relay to the centralized FLISR algorithm. When network connectivity is available, switch status messages are transmitted to the centralized algorithm via MQTT (Message Queuing Telemetry Transport) a lightweight, bi-directional communication protocol designed for resource constrained devices \cite{7123563}. To enable this interaction, an MQTT broker is deployed on the target system (e.g., DSO systems) which connects to the centralized FLISR algorithm, with the edge device exposing an outward MQTT interface. Control messages generated by the centralized FLISR algorithm are captured by the edge device and relayed to the RTU. In the event that network communication is unavailable, the edge device will leverage the local FLISR algorithm and pass the generated control message to the RTU directly through the IEC 104 connection, outlined in \hyperref[fig:rtu_interaction]{Figure \ref{fig:rtu_interaction}}.

 Switch status and control messages to define actions to take (i.e., opening/closing of the circuit breaker) contain a number of data points, an example is given below in JSON format.

\begin{lstlisting}[caption={Switch Example Status Message},captionpos=b]
{
    "LinkAdd": 1, "ASDU_CA": 51908,
    "TypeID": 30, "IOA": 4, "Value": 
    "0(OFF)"
}
\end{lstlisting}

 The properties in this data are, ``ASDU\_CA'' which is the identifier used by the DSO SCADA system to link the switch to its SCADA identifier, ``IOA'' (Information Object Address) which defines the specific input/output referenced on the switch, for example, the status of the power supply, and finally ``Value'' with, ``0(OFF)'' mapping to ``False'', with a value of ``1(ON)'' mapping to ``True''. 

As a general rule, any switch which gives an FPI and LVI signal, due to physical damage, should be opened, and therefore the damaged line or lines are no longer ``live''. The FPI and LVI message format for both a LBFM and Recloser switch are given below in addition to the corresponding control message to open or close the circuit breaker.

\begin{itemize}
  \item \textbf{LBFM switch FPI, LVI \& control structure}
\end{itemize}

 The FPI value for a LBFM switch map to IOA 4, meaning a ``Fault Current'' has been detected.

\begin{lstlisting}[caption={Switch fault current status},captionpos=b]
{
    "LinkAdd": 1, 
    "ASDU_CA": 51908,
    "TypeID": 30, 
    "IOA": 4, 
    "Value": "1(ON)"
}
\end{lstlisting}

 The LVI value maps to IOA 2, which specifies ``AC Supply Fail''.

\begin{lstlisting}[caption={Switch AC supply fail status},captionpos=b]
{
    "LinkAdd": 1, 
    "ASDU_CA": 51908, 
    "TypeID": 30, 
    "IOA": 2, 
    "Value": "1(ON)"
}
\end{lstlisting}

 An IOA value of 8 is used to open (``1(ON)'') or close (''0(OFF)'') the circuit breaker.

\begin{lstlisting}[caption={Switch C/B control instruction},captionpos=b]
{
    "LinkAdd": 1, 
    "ASDU_CA": 51908, 
    "TypeID": 30, 
    "IOA": 8, 
    "Value": "1(ON)"
}
\end{lstlisting}

\begin{itemize}
  \item \textbf{Recloser switch FPI, LVI \& control structure}
\end{itemize}

 The FPI value for a Recloser switch maps to an IOA of 27, specifying a ``End of Protection Sequence''.

\begin{lstlisting}[caption={Recloser End of Protection Sequence status},captionpos=b]
{
    "LinkAdd": 1,
    "ASDU_CA": 23089, 
    "TypeID": 30, 
    "IOA": 27, 
    "Value": "1(ON)"
}
\end{lstlisting}

 The LVI value maps to IOA 34 ``Controller Locked Out''.

\begin{lstlisting}[caption={Recloser Controller Locked Out status},captionpos=b]
{
    "LinkAdd": 1, 
    "ASDU_CA": 23089, 
    "TypeID": 30, 
    "IOA": 34, 
    "Value": "1(ON)"
}
\end{lstlisting}

 Finally, IOA 4096 is utilised to open (``1(ON)'') or close (''0(OFF)'') the circuit breaker.

\begin{lstlisting}[caption={Recloser C/B control instruction},captionpos=b]
{
    "LinkAdd": 1, 
    "ASDU_CA": 23089, 
    "TypeID": 30, 
    "IOA": 4096, 
    "Value": "1(ON)"
}
\end{lstlisting}

Based on what is received (FPI and/or LVI) and the characteristics of the pole, an appropriate control message is generated by FLISR and returned for a particular on pole circuit breaker. Both the centralized and distributed FLISR algorithms are implemented as lightweight containerised software services. Both algorithms are defined in the Go language, due to its suitability for building lightweight micro-services, performance and efficiency. A study by \textit{Dymora et al.} remarked that for processing small amounts of data, Go surpassed comparable languages in terms of performance, CPU and memory utilisation \cite{app10238521}, making it an ideal candidate for implementation of the FLISR algorithms where reaction speed is paramount.

% section methods (end)

%% file: Sections/4_experiments.tex
\section{Experiments}
\label{sec:experiments}

The simulation platform serves as an apparatus for demonstrating, testing and validating the FLISR algorithm outputs, in addition to bridging the gap between the research and industry by providing an intuitive tool to DSO personnel.
To ensure that the testing is as realistic as possible, the simulation platform implements interfaces to interact with the centralized and distributed FLISR algorithms that mimic those in the field. Furthermore, in order to evaluate the effects of network latency, the centralized algorithm and attached communication interfaces are deployed to a remote server, whereas the distributed FLISR algorithm operates locally, as would be the case in the field.

\begin{figure}[ht]
  \centering
  \includegraphics[width=0.4\textwidth]{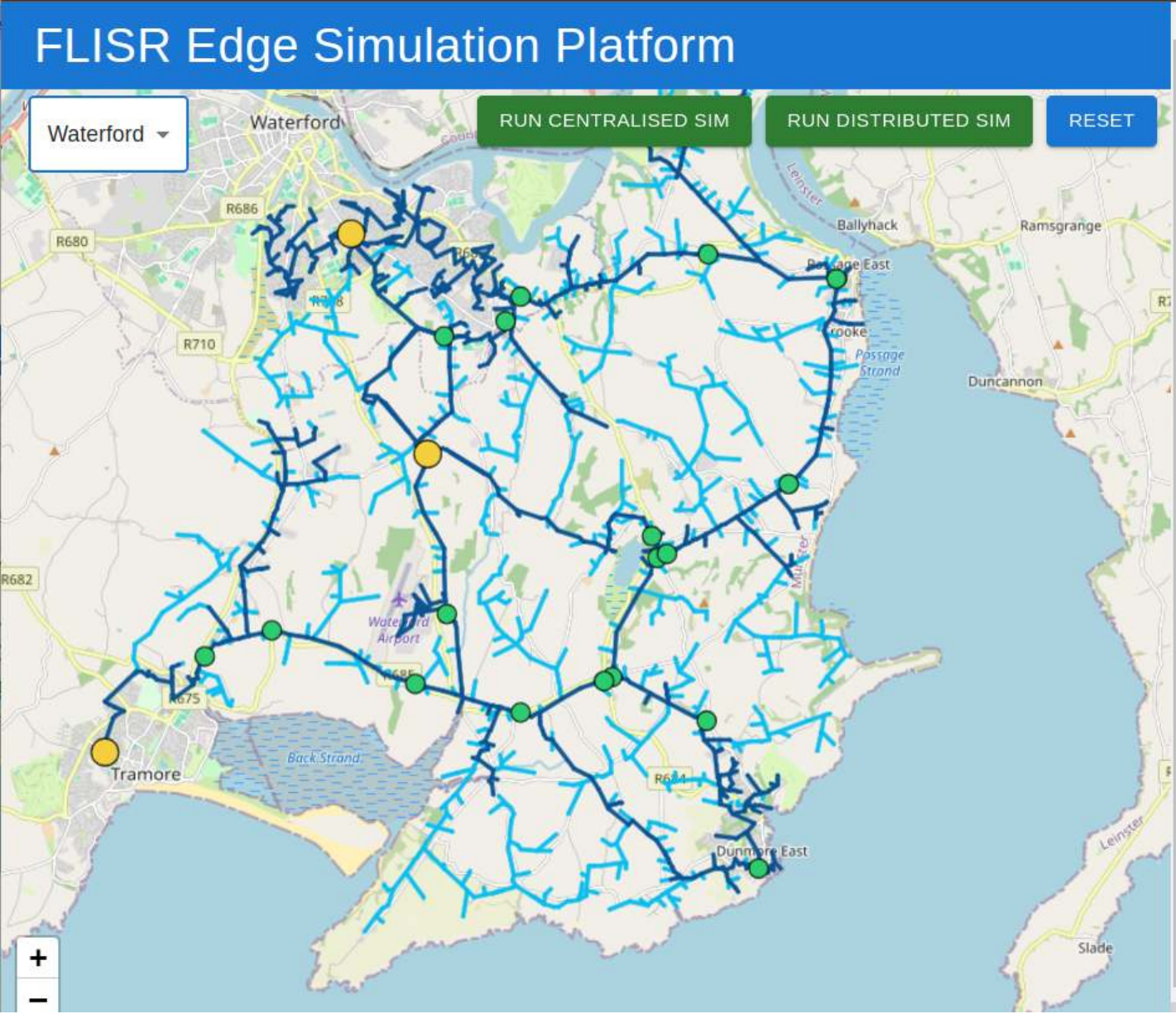}
  \caption{\label{fig:platform_ui}FLISR Edge simulation platform UI}
\end{figure}

The interface of the network simulation platform is accessed by the user via an internet browser, and presents the three trial sites overlaid on a geographical map populated with single and three phase lines, substations and on pole circuit breakers as presented in \hyperref[fig:platform_ui]{Figure \ref{fig:platform_ui}}.

\begin{figure}[ht]
  \centering
  \includegraphics[width=0.4\textwidth]{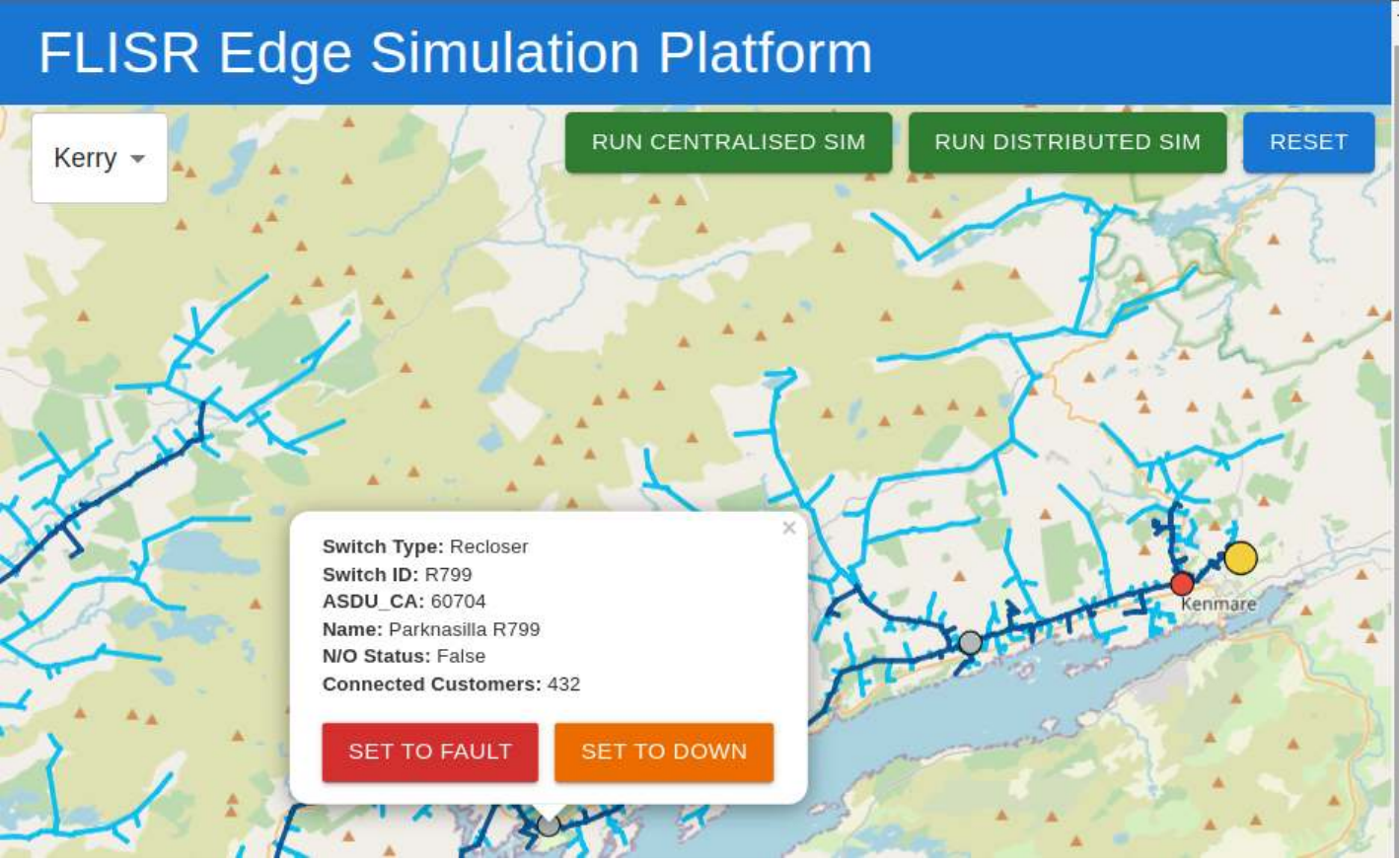}
  \caption{\label{fig:test_ui}FLISR Edge UI test case example}
\end{figure}

 Each switch asset on the map is a clickable object, at which point the user is presented with a box containing the properties of the asset, such as whether it is configured as a normally open point and the amount of connected customers. In addition, each switch has the option to either set it to faulted or down status. By selecting a faulted status the simulation platform generates an FPI and LVI status message which corresponds to the switches type (LBFM or Recloser), setting the switch to down status, generates an LVI status message. Faulted switches are visualised as red markers, whereas down switches are grayed out, and in this way the user can intuitively build a fault case to test as illustrated in \hyperref[fig:test_ui]{Figure \ref{fig:test_ui}}.

 Test cases can be run in either centralized or distributed FLISR mode. Depending on the option chosen, the FLISR Edge software service will generate the corresponding FPI and LVI status messages and pass them to the relevant FLISR algorithm. For the centralized FLISR algorithm, the generated FPI and LVI messages are passed upstream via MQTT to the remote server, with each generated control message is passed downstream based on the assets unique ASDU indentifier.

\begin{figure}[ht]
  \centering
  \includegraphics[width=0.5\textwidth]{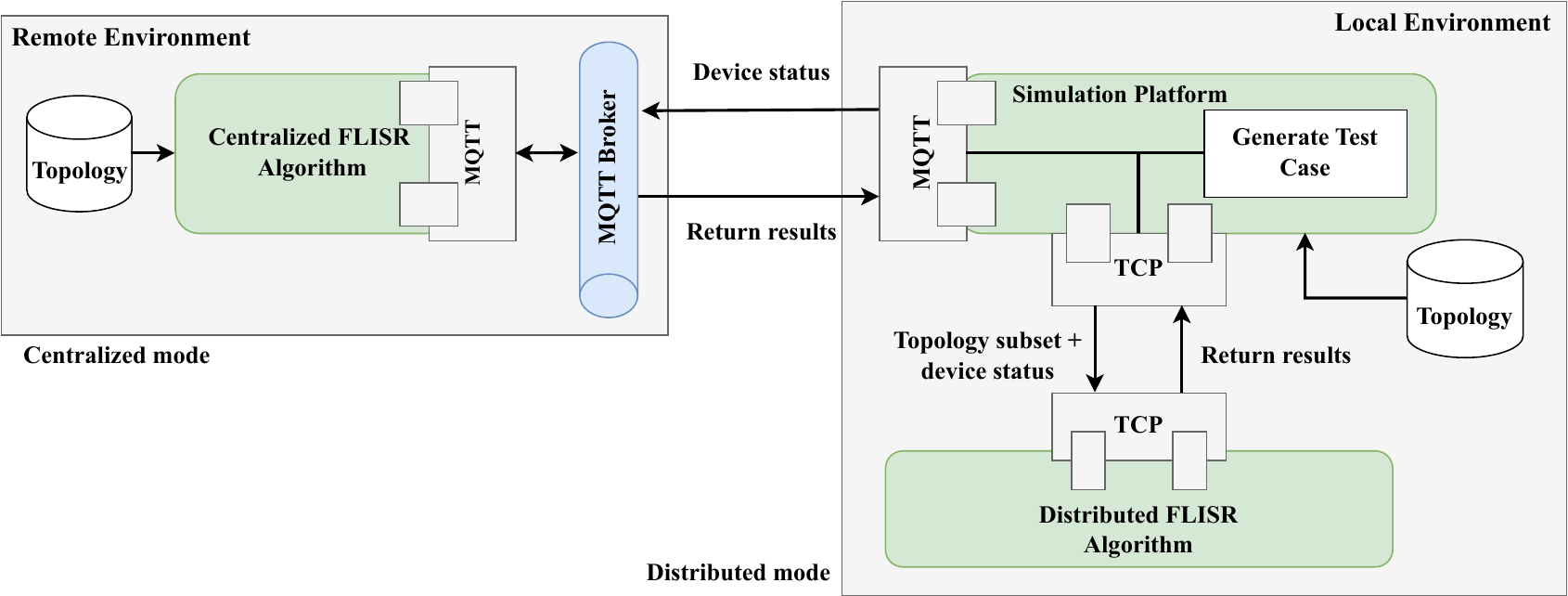}
  \caption{\label{fig:sim_interaction}Simulation platform interfaces \& interaction}
\end{figure}

 When running in distributed mode, the FLISR Edge software passes the FPI \& LVI messages to the local FLISR algorithm. These are passed via TCP, in addition to the properties of the pole. This mechanism mimics how the operation would occur when the distributed FLISR algorithm is deployed to an edge device and installed in the field, in order to realistically simulate fault scenarios. This interaction of the simulation platform with the FLISR algorithms is described in \hyperref[fig:sim_interaction]{Figure \ref{fig:sim_interaction}}.

\begin{figure}[ht]
  \centering
  \includegraphics[width=0.4\textwidth]{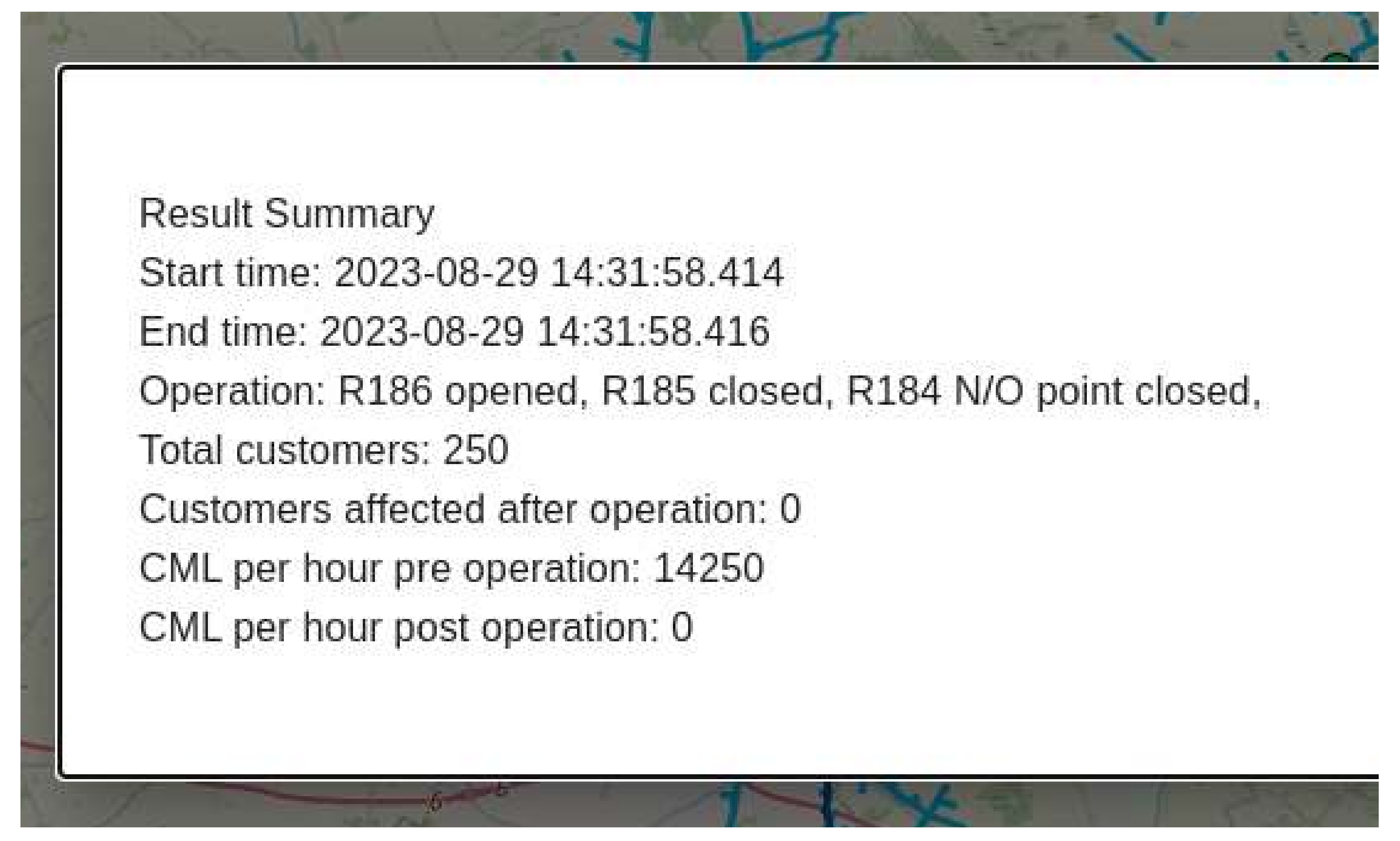}
  \caption{\label{fig:ui_result}Simulation result (user interface)}
\end{figure}

 Control messages generated by the FLISR algorithms are captured by the simulation platform, and the results are calculated. Result metrics include time of operation, an overview of assets affected as part of the test case, the control operations taken as a result of the FLISR algorithms output and the total number of affected customers before and after FLISR operation. Based on this figure a calculation of CML per hour before and after the FLISR operation is calculated. These generated results are returned to the user via the FLISR Edge network simulation platform UI as presented in \hyperref[fig:ui_result]{Figure \ref{fig:ui_result}}. These results written to a CSV file, which is updated with each test case run. An example output of the CSV report is presented in \hyperref[table:1]{Table \ref{table:1}}.

\begin{table*}[ht]
\vspace{0.5cm}
    \centering
    \caption{\label{table:1}Distributed FLISR algorithm sample results CSV}
    \begin{adjustbox}{width=\textwidth}
    \begin{tabular}{lllllrrrrp{5cm}}
    \toprule
        Start time & End time & Faulted devices & Down devices & Operation & Affected customers (pre) & Affected customers (post) & CML per hour (pre) & CML per hour (post) \\ 
        \midrule
        2023-08-25 08:11:37.300 & 2023-08-25 08:11:37.303 & S513 & S512, S716, S514, S700 & S513 opened, S512 closed, S716 closed, S514 N/O point closed, S700 N/O point closed & 627 & 0 & 35739 & 0 \\ 
        2023-08-29 14:31:58.414 & 2023-08-29 14:31:58.416 & R186 & R185, R184 & R186 opened, R185 closed, R184 N/O point closed & 250 & 0 & 14250 & 0 \\
        2023-08-29 14:35:55.852 & 2023-08-29 14:35:55.854 & S507, S508 & R518, S509, S697 & S507 opened, S508 opened, R518 closed, S509 closed, S697 N/O point closed & 1370 & 171 & 78090 & 9747 \\ 
        2023-08-29 14:36:35.252 & 2023-08-29 14:36:35.255 & R517, S513, S512 & S716, S514, S700 & R517 opened, S513 opened, S512 opened, S716 closed, S514 N/O point closed, S700 N/O point closed & 784 & 278 & 44688 & 15846 \\ 
    \bottomrule
    \end{tabular}
    \end{adjustbox}
\end{table*}

% section experiments (end)

%% file: Sections/5_results.tex
\section{Results \& Future Work}
\label{sec:results}

%\subsection{Research findings}
%\label{sec:findings}

\begin{itemize}
  \item \textbf{Effectiveness of the FLISR Solution}
\end{itemize}

The experiments reveal the effectiveness of FLISR in reducing fault KPIs depends on specific circumstances.  Networks with fewer faulted switch devices and less complex structures see less significant improvements. FLISR impact also lessens when a network experiences numerous simultaneous faults. In a Waterford grid simulation, a single faulted switch initially caused 1370 affected customers and 78090 CML/hour. FLISR reduced this to 171 customers and 9747 CML/hour. However, simulating four faulted switches on the same grid segment resulted in 844 affected customers and 48108 CML/hour, even after FLISR intervention.

\begin{itemize}
  \item \textbf{Centralized vs. Distributed FLISR Algorithms}
\end{itemize}

Comprehensive testing found the operational steps generated by the algorithms are identical, with no observable difference in terms of the steps taken to resolve a given fault event. In practice, this is a desirable outcome as when applied to the field, divergent control operations generated by each algorithm may suggest that the algorithms require refactoring. Furthermore, it is not definitive at this stage that identical results will be given for every situation, as only three networks have been tested in a simulated environment. With further testing and validation other factors and distribution network features may be found which impact the FLISR algorithm logic.

A difference observed between the centralized and distributed FLISR algorithms is the operational time. As the centralized algorithm requires a network connection to ingest and output generated control signals, network latency becomes a factor. To better ascertain the impact of network latency on the centralized FLISR operation, an identical test case was simulated twenty times on a 5G network, 4G LTE, 3G and 2G, with the average time calculated for each network type as presented in \hyperref[table:2]{Table \ref{table:2}}.

\begin{table}[h]
    \centering
    \small
    \caption{\label{table:2}Average centralized FLISR operational speed}
    \renewcommand{\arraystretch}{0.5} % adjust the spacing between rows
    \begin{tabular}{@{}l r@{}}
    \toprule
     Network type & Average speed (ms)  \\
     \midrule
     5G & 212.8  \\
     4G LTE & 260.4  \\
     3G & 267.1  \\
     2G & 1049.6 \\
    \bottomrule
  \end{tabular}
\end{table}

The average operational speed when using low bandwidth networks such as 2G is quite high, this is significant as when applied to the field, the most commonly available networks available are likely to be 3G and 2G, particularly in rural areas. When compared to the distributed FLISR algorithm where network latency is not a consideration, for the tests conducted the average speed of operation is at or near real-time at two milliseconds.

\begin{itemize}
  \item \textbf{FLISR Solution Adaptability}
\end{itemize}

 As the findings gathered from the results above have shown, the FLISR Edge solution produces control signals for each of the test cases simulated on the three trial distribution networks. As stated previously, the effectiveness of the solution depends on the network configuration and the level of simulated damage to the grid, for simpler networks such as Kerry and Mullingar, the effectiveness of the solution can be more limited compared to a radial network such as Waterford in terms of reducing the number of affected customers and CML accumulated.
However, another aspect of quickly isolating faulted networks is one of safety, where downed lines which are still live pose a safety threat to citizens and therefore the immediate operation enabled by the FLISR Edge solution would assist in alleviating this danger.

% subsection findings (end)

\subsection{Relevance to the DSO \& future work}
\label{sec:relevance}

Ireland's energy regulator financially penalizes DSOs for excessive service interruptions, creating a strong incentive to implement solutions like FLISR Edge. As climate change intensifies, weather-related damage and faults will likely increase, making such solutions even more valuable to both DSOs and consumers. FLISR Edge offers a cost-effective, fault-tolerant approach that leverages edge computing for independent network operation and significant reductions in customer outages.

\textbf{\emph{Bridging the Gap Between Research and Industry}} - Our network simulation platform is crucial for demonstrating the capabilities of FLISR Edge and bridging research with industry applications. By engaging DSO personnel in testing, validation, and feedback, we can continually improve the solution. This collaboration will not only enhance FLISR Edge but also serve as a valuable template for developing future smart energy and edge computing proposals. DSOs can leverage the platform to assess the potential impact of FLISR Edge on their specific networks, identify areas for optimization, and gain insights into how it can integrate seamlessly with their existing infrastructure.

\textbf{\emph{Continuous Improvement and Future Developments}} - While the current simulation platform lacks detailed energy flow modeling (which could be enhanced with SCADA data), it offers a solid foundation for future development. One promising avenue is exploring predictive analysis to identify transient faults. This would enable DSOs to proactively address issues (e.g., overgrown tree lines), potentially preventing sustained faults and further outages.  Additionally, the platform can be enhanced to incorporate real-time weather data, allowing for simulations that more accurately reflect the impact of various weather conditions on grid performance. This would provide DSOs with even more valuable insights into how FLISR Edge can improve grid resilience under a wider range of environmental stresses.

% subsection relevance (end)

% section results (end)

%% file: Sections/6_conclusion.tex
\section{Conclusion}
\label{sec:conclusion}

The primary contribution of this paper is to demonstrate the critical role of the energy network in modern society and how climate change-related weather events disrupt energy supply, affecting consumers and incurring financial penalties for DSOs. FLISR emerges as a powerful solution to mitigate these impacts. Our research builds upon existing work, leveraging modern software, communication technologies, and DSO data streams to create a fault-tolerant FLISR solution. By combining remote and edge computing capabilities, we automate fault detection and resolution to minimize downtime and its financial consequences. Our simulation platform, with its intuitive interface, bridges the academic-industry gap. It allows us to demonstrate the FLISR Edge solution to DSO personnel, gather valuable feedback for continual improvement, and explore commercialization. This work highlights the potential of such tools to extend beyond FLISR. Our approach can inform the development of ancillary services for grid stability, industrial and domestic energy management systems, and intelligent energy flexibility solutions. These applications are crucial for addressing the challenges posed by increased renewable energy sources, distributed generation, and the rise of prosumers.

% section conclusion (end)